\documentclass[aps,pra,twocolumn,showpacs,groupedaddress]{revtex4-1}

\usepackage{amsfonts,mathrsfs,amsmath,amsthm,amssymb,bbold}
\usepackage{epsfig}
\usepackage{bm}
\usepackage{tabulary}
\usepackage{color}
\usepackage[dvipsnames]{xcolor}
\usepackage{hyperref}

\begin{document}  
\title {\bf Systematic errors in direct state measurements\\
with quantum controlled measurements} 

\author{ Le Bin Ho}
\thanks{Electronic address: binho@kindai.ac.jp}

\affiliation{Department of Physics, Kindai University, Higashi-Osaka, 577-8502, Japan}
\affiliation{Ho Chi Minh City Institute of Physics, VAST, Ho Chi Minh City, Vietnam}

\date{\today}

\begin{abstract}

Von Neumann measurement framework
describes a dynamic interaction between 
a target system and a probe. 
In contrast, a quantum controlled
measurement framework uses a 
qubit probe to control the actions of different operators on
the target system, and convenient for  
establishing universal quantum computation.
In this work, we use a quantum controlled measurement framework
for measuring quantum states directly. 
We introduce two types of 
the quantum controlled measurement framework 
and investigate the systematic error 
(the bias between the true value 
and the estimated values) that caused by these types. 
We numerically investigate the systematic errors, 
evaluate the confidence region,
and investigate the effect of experimental noise
that arises from the imperfect detection.
Our analysis has important applications 
in direct quantum state tomography.

\end{abstract}
%
%

\maketitle

\section{Introduction}\label{sec_i}
Quantum state tomography (QST) is a process of 
getting the information of
quantum states from measurement data
\cite{Paris2004}. 
Under the tremendous growth in 
quantum technologies, 
QST is of vital importance for benchmarking, 
experimentally validating quantum devices, 
and establishing new quantum technologies. 
Therefore, it induces a critical demand to 
develop effective schemes for the QST and 
evaluate their efficiency, 
including innovation, practically realizable, 
and significance.

Typically, a standard quantum state tomography (sQST) includes 
(i) the measurements of multiple copies of a system 
in a complete set of noncommuting observables, and 
(ii) the reconstruction of the most likely quantum state 
from the measured data set using efficient algorithms
such as linear inversion, maximum-likelihood, least squares estimations, etc
\cite{PhysRevA.61.010304}.
In the past years, many endeavors have been devoted 
to rising the efficiency of the sQST
\cite{Kosaka2009,Vanner2013,PhysRevLett.116.230501,PhysRevA.93.052105}.
It is, however, particularly challenging to 
apply for high-dimensional systems 
because it requires dramatically increasing 
measurement cost 
(a full state tomography of a $d$-dimensional system 
requires at least $d^2-1$ different measurements)
and thus, consumes substantial calculation time. 
Numerous practical works have been dedicated 
to reducing the number of measurements, 
emergent with compressed sensing 
\cite{PhysRevLett.105.150401,Flammia_2012}, 
reduced density matrix \cite{PhysRevLett.118.020401},
and adaptive quantum tomography
\cite{PhysRevA.98.032330}.
So far, recent achievements on 
dynamics enhancement and dynamics control of
entangled systems have been reported
\cite{doi:10.1142/S0219749915500343,Ateto2017},
which pave the way for studying tomography of 
entangled systems.

Separate from the sQST, 
a direct state measurement (DSM) method
allows for measuring the wave functions directly 
\cite{Lundeen2011}.
It was originally proposed by Lundeen et al. 
based on the evidence that 
the amplitudes of the wave functions are 
proportional to weak values \cite{Lundeen2011}.
The DSM has more experimental merit than the sQST 
because it is straightforward, simple, versatile, 
required only local measurements
\cite{PhysRevLett.108.070402},
and also be able to apply for large systems 
\cite{Shi:15,PhysRevLett.113.090402,Malik2014,Bolduc2016,PhysRevA.98.023854}. 
It was enormously extended to general mixed states  
\cite{PhysRevLett.108.070402,PhysRevLett.117.120401,PhysRevLett.121.230501},
phase-space distributions 
\cite{PhysRevLett.108.070402,PhysRevA.86.052110,de_Gosson_2012,PhysRevA.98.023854},
enlarged Hilbert space \cite{HO2019289},
and nonlocal entangled states, recently
\cite{PhysRevLett.123.150402}.

There are numerous reports on the 
improving measurement precision in the DSM
using strong interaction measurements
\cite{PhysRevLett.116.040502,PhysRevLett.118.010402,PhysRevLett.121.230501},
compressive sensing 
\cite{PhysRevA.97.032120},
enlarged Hilbert space  
\cite{HO2019289}, and also
continuous probe state
\cite{ZHU2017283}.
Besides, examining the novelty, efficacy,
and significance of the DSM had been reported 
\cite{PhysRevA.92.062133}.
The statistical error estimation 
\cite{Sainz_2016}
and the protection of the systematic errors 
when the probe undergoes decoherence 
\cite{PhysRevA.94.012329}
have also been focused recently.

Measurements used in the DSM typically 
described by a von Neumann model,
which requires a dynamical interaction 
between a target system and a probe
\cite{vonNeumann}, see also
\cite{PhysRevA.98.042112,Botero2018,DENKMAYR2018339}
for strong interaction measurements of weak values.
Recently, Ogawa et al.~\cite{Ogawa_2019}, 
however, has proposed a measurement framework
of ``probe-controlled-system transformation"
that without using the von Neumann measurement.
That framework actually can be derived from the
von Neumann measurement,
and thus they are operationally equivalent. 
In that framework, a target system interacts 
with a control qubit probe via the actions
of different operators controlled by the qubit probe
and can be seen as a ``quantum controlled 
measurement"
(see Sec. \ref{sec_iiA} below.)

Previously, Hofmann \cite{F_Hofmann_2014} has also
developed a framework of the quantum controlled 
measurement
where the control qubit probe controls the operations
of zero interaction (identity operator) 
or fully projective measurement on the target system.
This method is a kind of quantum controlled gate,
and the same procedure has been experimentally verified 
for measuring weak values
\cite{PhysRevLett.108.070402,PhysRevLett.112.070405}.
 
An excellent feature of the quantum controlled 
measurement is its applicability
of the cyclic transformation property,
which paves the way for using a scan-free method \cite{Shi:15}
(see discussion on Sec. \ref{sec_iv} below.)
A scan-free method is that all the data after 
a post-selection process will be kept and used for estimating
quantum states.
Therefore, it will help to improve the measurement 
precision in the DSM.
 
In this work, we study the systematic 
errors in the DSM 
caused by different operational types in
quantum controlled measurements. 
By definition, systematic error is
a bias between the true value and the estimated values.
We consider two types of operational interaction, 
i.e., type-I and type-II, correspond to 
Hofmann's and Ogawa's frameworks, respectively.
We first analyze the systematic errors 
caused by these types. Then, we also 
compare their efficiency by evaluating 
the confidence region. We finally 
investigate the systematic errors under 
the imperfect detection noise,
i.e., the noise that
arises from the imperfection of the measuring detectors. 
In these calculations, we analyze the fidelity, 
a figure of merit that is obtained from 
Monte Carlo simulations.

We emphasize that type-II corresponds 
to an arbitrary strong interaction DSM. 
It thus represents the prevailing evolutions 
of the DSM. Whereas, the performance of 
type-I on the DSM has not been known yet. 
Therefore, it is essential to investigate and 
compare the efficiency of the DSM via these 
two types of operational quantum controlled 
measurement.

The structure of this paper is organized as follows. 
Section \ref{sec_ii} introduces the measurement schemes 
of the DSM using two types of operational 
quantum controlled measurement. 
Section \ref{sec_iii} presents the main numerical results 
of the systematic errors, including the investigation 
of the confidence region and the effect of noise. 
Section \ref{sec_iv} is devoted to a discussion.
The paper concludes with a summary in Sec. \ref{sec_v}.

\section{Direct state measurement
with quantum controlled interaction}\label{sec_ii}

\subsection{Quantum controlled interaction}\label{sec_iiA}
Let us first introduce a general form of quantum controlled 
measurement framework \cite{F_Hofmann_2014,Ogawa_2019}.
In this framework, a target system 
is controlled by
a control qubit probe. 
The interaction between the target system and
the control qubit probe is given by
\begin{align}\label{eq:inter}
\bm{U} = \bm U_1\otimes|0\rangle\langle 0|
+ \bm U_2 \otimes|1\rangle\langle 1|,
\end{align}
where $\bm U_i, (i = 1, 2)$ are two operators operate 
on the target system, 
and $|0\rangle, |1\rangle$ are two bases of the control qubit probe.
Such operators can be implemented in experiments if
their absolute eigenvalues 
are smaller or equal to one \cite{Ogawa_2019}.

We next apply the quantum controlled measurement framework
to measure the quantum state directly. 
We consider two types of operational interaction
in (\ref{eq:inter}) as following. 

\subsection{Type-I operational interaction}\label{type-I}

\begin{figure} [t]
\centering
\includegraphics[width=8.4cm]{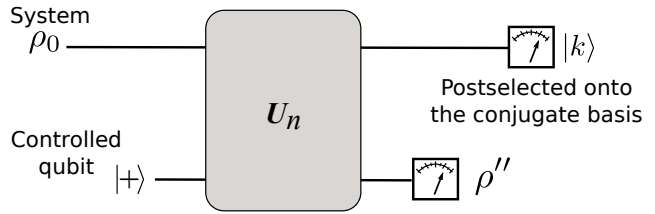}
\caption{
(Color online.) 
Scheme of the direct state measurements (DSM).
A control qubit probe is coupled to a target system 
(in $d$-dimensional Hilbert space)
via a unitary $\bm U_n, n \in [0, d-1]$.
The target system is then postselected onto the 
conjugate basis $\{k\}$, while the control qubit probe is measured 
on different bases, i.e., $\{ 0, 1\}, \{+,-\}, \{L,R\}$
to reproduce the  target system state $\rho_0$.
}
\label{fig1}
\end{figure}

We consider a measurement scheme 
between a target system
and a control qubit probe as 
schematically shown in Fig.~\ref{fig1}. 
The target system is initially given 
in the density matrix $\rho_0$, 
\begin{align}\label{eq:rho0}
\rho_0 = \sum_{n,m=0}^{d-1}
\rho_{nm}|n\rangle\langle m|,\
\text{ with } \rho_{nm} = 
\langle n|\rho_0|m\rangle,
\end{align}
an unknown state needed to be estimated,
$d$ is the dimension of the target system.
The control qubit probe is initially prepared in the state $|+\rangle$,
i.e., $|+\rangle = \bigl(|0\rangle + |1\rangle\bigr)/\sqrt{2}$.
The initial joint state becomes
\begin{align}\label{eq:rho_B}
\rho = \rho_0\otimes |+\rangle\langle+|.
\end{align}
Let us consider the interaction is an invert 
quantum controlled gate
which we name as type-I operational interaction:
\begin{align}\label{eq:U_B}
\bm{U}_n = \bm{I}_{\rm s} \otimes |0\rangle\langle 0|
+ |n\rangle\langle n|\otimes |1\rangle\langle 1|,
\end{align}
where $\bm{I}_{\rm s}$ is a $d$-dimensional 
identity matrix in the target system space. 
Note that this interaction is similar to Hofmann's and
can be implemented in optics
where the control qubit probe is a polarized single photon
\cite{F_Hofmann_2014}. 
This kind of implementation has been reported in 
\cite{PhysRevLett.108.070402,PhysRevLett.112.070405}.

After the interaction, the joint state becomes
\begin{align}\label{eq:rho'}
\rho' = \bm{U}_n \rho \bm{U}_n^\dagger.
\end{align}
The target system is then postselected onto a conjugate basis
$|k\rangle = \frac{1}{\sqrt{d}}\sum_{m=0}^{d-1} e^{i2\pi mk/d}|m\rangle$,
while the remaining control qubit state is given by
\begin{align}\label{eq:rho''}
\rho'' = \langle k|\rho'|k\rangle = 
\begin{pmatrix}
    \rho''_{00}(n,k) & \rho''_{01}(n,k)\\
    \rho''_{10}(n,k) & \rho''_{11}(n,k)
\end{pmatrix}.
\end{align}
Explicitly, we have   
\begin{align}\label{eq:rhoij}
\rho''_{00}(n,k) &= \frac{1}{2d}\sum_{n,m=0}^{d-1} e^{\frac{i2\pi(m-n)k}{d}}\rho_{nm};\\
\rho''_{10}(n,k) &= \frac{1}{2d}\sum_{m=0}^{d-1} e^{\frac{i2\pi(m-n)k}{d}}\rho_{nm};\\
\rho''_{01}(n,k) &=\rho''_{10}(n,k)^*; \text{ and } \\
\rho''_{11}(n,k) &= \frac{1}{2d}\rho_{nn}.
\end{align}

Using the Fourier transformation, we obtain
\begin{align}\label{eq:rho_nm_I}
\rho_{nm} \propto 2d \sum_k e^{\frac{i2\pi(n-m)k}{d}}\rho''_{10}(n,k).
\end{align}
Then, the element $\rho_{nm}$ is calculated
from the measurement result of $\rho''_{10}(n,k)$ in the control qubit probe. 
To obtain $\rho''_{10}(n,k)$, 
we measure the control qubit probe in different bases:
\begin{align}
\rho''_{10}(n,k) &= \dfrac{1}{2}
\Bigl[\bigl(P_+ -P_-\bigr) +i\bigl(P_{L} -P_{R}\bigr)\Bigr],\label{eq:rho10_I}
\end{align}
where $P_j = {\rm tr} [|j\rangle\langle j|\rho'']$ is the 
probability when measuring the control qubit probe
in different bases;
where $|j\rangle \in \{ |0\rangle, |1\rangle\}, 
\{|+\rangle, |-\rangle\}, \{|L\rangle, |R\rangle\}$,
where 
$|\pm\rangle = \frac{1}{\sqrt{2}}\bigl(|0\rangle\pm|1\rangle\bigr),
|L\rangle = \frac{1}{\sqrt{2}}\bigl(|0\rangle + i|1\rangle\bigr), 
|R\rangle = \frac{1}{\sqrt{2}}\bigl(|0\rangle - i|1\rangle\bigr)$.

\subsection{Type-II operational interaction}\label{type-II}
We consider an arbitrary coupling between the target system 
and the control qubit probe. 
The initial joint state is the same as \eqref{eq:rho_B} in Sec. \ref{type-I}.
The interaction is given by
\begin{align}\label{eq:U_B_II}
\bm{U}_n = \Bigl(\bm{I}_{\rm s}-
\varepsilon_\theta |n\rangle\langle n|\Bigr) 
\otimes|0\rangle\langle 0|
+ \sin\theta |n\rangle\langle n|\otimes |1\rangle\langle 1|,
\end{align}
which we name as type-II operational interaction.
Here $\varepsilon_\theta \equiv 2\sin^2\frac{\theta}{2}$,
and $\theta$ is the coupling strength. 
For $\theta \ll 1$, the measurement is said to be weak.
For $\theta = \pi/2$, the measurement is strong, 
while $\theta < \pi/2$ corresponds to 
an arbitrary strength measurement. 
This type is equivalent to 
an arbitrary strong von Neumann measurement framework
\cite{PhysRevLett.116.040502,PhysRevLett.118.010402,PhysRevA.89.022122,PhysRevA.84.052107,CHEN20173161,PhysRevA.93.032128,PhysRevA.93.062304},
such that, as we can see
in the von Neumann measurement, 
a unitary interaction is given by 
\begin{align}\label{eq:interaction}
\notag\bm{U}_n^{\rm vNm} &= e^{-i\theta|n\rangle\langle n|\otimes\sigma_y}\\
& = \bm{I}_{\rm sp} -|n\rangle\langle n|\otimes
\bigl[(1-\cos\theta)\bm{I}_{\rm p}+i\sin\theta\sigma_y\bigr],
\end{align}
where $\bm{I}_{\rm sp} = \bm{I}_{\rm s}\otimes \bm{I}_{\rm p}$;
$\bm{I}_{\rm s}, \bm{I}_{\rm p}$ are identity matrices in the  target system
and control qubit probe, respectively. 
vNm stands for ``von Neumann measurement."
The action of $\bm{U}_n^{\rm vNm}$ on the control qubit probe 
initially prepared in state $|0\rangle$ leads to
\begin{align}
\bigl(\bm{I}_{\rm s}-\varepsilon_\theta|n\rangle\langle n|\bigr)
\otimes |0\rangle\langle 0| + \sin\theta |n\rangle\langle n|
\otimes |1\rangle\langle 1|,
\end{align}
which is the same as \eqref{eq:U_B_II}
above.
This type of measurement covers  all of the 
versions of DSM 
from strong \cite{PhysRevLett.116.040502,PhysRevLett.118.010402} 
to weak interaction \cite{PhysRevA.89.022122,PhysRevA.84.052107,CHEN20173161} 
by changing the coupling strength. 
This method also includes the coupling-deformed-pointer method,
which is an arbitrary strong interaction 
\cite{PhysRevA.93.032128,PhysRevA.93.062304}.

Next, after the interaction  given in \eqref{eq:U_B_II} and 
the postselection onto the conjugate basis $|k\rangle$,
the final control qubit state ($\rho''$) is given
as in \eqref{eq:rho''}.
Then, we have
\begin{align}\label{eq:rhoij-II}
\notag\rho''_{00}(n,k) &= \frac{1}{2d}
\Bigl[\sum_{n,m=0}^{d-1} e^{\frac{i2\pi(m-n)k}{d}}\rho_{nm}\\
&-\varepsilon_\theta
\Bigl(\sum_{m=0}^{d-1} e^{\frac{i2\pi(m-n)k}{d}}\rho_{nm}+c.c\Bigr)
+\varepsilon_\theta^2\ \rho_{nn}\Bigr];\\
\rho''_{10}(n,k) &= \frac{\sin\theta}{2d}
\Bigl[\sum_{m=0}^{d-1} e^{\frac{i2\pi(m-n)k}{d}}\rho_{nm}
+\varepsilon_\theta\rho_{nn}\Bigr];\\
\rho''_{01}(n,k) &=\rho''_{10}(n,k)^*; \text{ and } \\
\rho''_{11}(n,k) &= \frac{1}{2d}\sin^2\theta\ \rho_{nn}.
\end{align}

It is now possible to calculate the 
element $\rho_{nm}$ by using 
the Fourier transformation: 
\begin{align}\label{eq:rho_nm_II}
\rho_{nm} \propto 2d\tan\frac{\theta}{2}\delta_{nm}\rho''_{11}(n,k) 
+\sum_k e^{\frac{i2\pi(n-m)k}{d}}\rho''_{10}(n,k),
\end{align}
where $\delta_{nm}$ is the Dirac delta. 
For a weak measurement (small $\theta$), 
the reconstructed state is given by
\begin{align}\label{eq:rho_nm_weak}
\rho_{nm}^{\rm W} \propto 
\sum_k e^{\frac{i2\pi(n-m)k}{d}}\rho''_{10}(n,k).
\end{align}

The same as Sec. \ref{type-I}, 
to obtain $\rho''_{10}(n,k)$ and $\rho''_{11}(n,k)$,
we measure the control qubit probe as follows:
\begin{align}
\rho''_{10}(n,k) &= \dfrac{1}{2}
\Bigl[\bigl(P_+ -P_-\bigr) +i\bigl(P_{L} 
-P_{R}\bigr)\Bigr],\ \text{and} \label{eq:rho10_II}\\
\rho''_{11}(n,k) &= P_1, \label{eq:rho11_II}
\end{align}
where $P_1 =  {\rm tr} [|1\rangle\langle 1|\rho'']$.
Equations (\ref{eq:rho_nm_II}, \ref{eq:rho_nm_weak}) 
are equivalent to those derived by 
von Neumann measurement in 
\cite{PhysRevLett.116.040502}.
%
This Neumann framework has been widely explored 
and investigated theoretically and experientially 
 \cite{PhysRevLett.116.040502,PhysRevLett.118.010402}. 

Such an element matrix 
$\rho_{nm}$ in
Eqs.~(\ref{eq:rho_nm_I}, \ref{eq:rho_nm_II}) 
is given without any approximation. However, we emphasize that its obtained value depends on the experiments and the methods (that we use to measure the  final control qubit state):
by using different types of equipment or different methods, we have different values of $\rho_{nm}$, which is the main leaven of systematic errors.
So far, there are various methods in the DSM,
including compressive sensing method 
\cite{PhysRevA.97.032120},
enlarged Hilbert space \cite{HO2019289}, 
and continuous probe state \cite{ZHU2017283}.
However, in this work, we restrict ourselves to 
quantum controlled measurement framework
since it includes the arbitrary strong interaction
and is experimentally realizable
\cite{PhysRevLett.118.010402,Ogawa_2019}.

{\section{Accuracy in the DSM}\label{sec_iii}}

\subsection{Random vs systematic errors}
To avoid any confusion, 
we define random and systematic errors as following.
Random error describes fluctuations around the true value that
caused by
 unknown and unpredictable changes
during the measurement process.
It is essentially unavoidable
but can be effectively reduced by repeating 
the measurement many times. 
Differently, systematic error is 
 a bias between 
the true value and the estimated value 
\cite{DURKEE2006191}. 
It is raised by 
inaccuracy of equipment 
or by diffident measurement methods (models),
 such as the offset error, the scale factor error.
For example, if you are using a thermometer that has not
been set to zero beforehand, there will have a systematic error
in measuring the temperature (offset error); 
or using a stretched-out measuring tape will cause
a scale-factor type systematic error.
In our case, the systematic error is caused
by  diffident types in the quantum controlled 
measurement framework.

\begin{figure} [t]
\centering
\includegraphics[width=8.2cm]{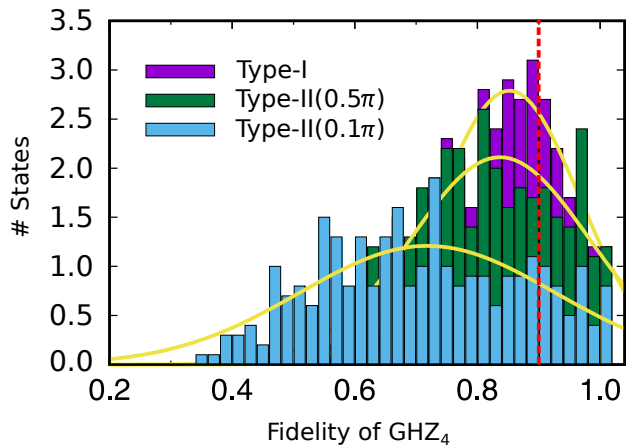}
\caption{
(Color online.) 
Histogram of the fidelities estimated from 500 
independent reconstructed states $|\rm {GHZ}_4\rangle$. 
For each state, $N_c = 400$ measurement repetitions
have been carried out. The red-dotted line is the reference fidelity $f_0$.
The average fidelities obtained from type-I (purple),
type-II($0.5\pi$) (green), 
and type-II($0.1\pi$) (cyan) 
are smaller than $f_0$, which called systematic errors.
Yellow curves are fit Gaussian. 
}
\label{fig2}
\end{figure} 

We numerically investigate 
the systematic errors in the DSM
 caused by two types of operational interaction 
in Sec. \ref{sec_ii}.
We first consider a four-qubit GHZ state 
$|{\rm GHZ_4}\rangle = 1/\sqrt{2} (|0000\rangle+|1111\rangle)$
 as a target state
(later, we also examine other common states such as W state and Dicke state.)
We also assume the state is mixed with small white noise
that $\rho_0 = (1-p) |{\rm GHZ_4}\rangle\langle{\rm GHZ_4}|
+ p \bm{I}_{\rm s}/16$.
We choose $p$ so that the reference fidelity 
$f_0 \equiv \langle{\rm GHZ_4}|\rho_0|{\rm GHZ_4}\rangle = 0.9$.
To reconstruct  the state, 
we perform the Monte Carlo simulations with the cumulative method 
\cite{PhysRevA.89.022122,HO2019289}. 
Our code can be found in \cite{Code}.
 To analyze the systematic errors, we compare the bias 
between the reference fidelity and average fidelities obtained 
from type-I and type-II as follows. 



Figure \ref{fig2} shows the histogram of the estimated 
fidelity $f(\rho_r) = \langle{\rm GHZ_4}|\rho_r|{\rm GHZ_4}\rangle$,
where $\rho_r$ is the estimated density state,
obtained from \eqref{eq:rho_nm_I} or \eqref{eq:rho_nm_II}.
Note that this fidelity is compared with the GHZ state,
so that its exact value is $f_0 = 0.9$, as we fit above. 
 If we compare $\rho_r$ with $\rho_0$
then the exact value of the fidelity should be one. 
The reason we shift the exact value
to 0.9 is that for the illustration purpose, as we can see from Fig.~\ref{fig2}. 
From the histogram, we can calculate the average fidelity
($f_{\rm ave}$),
which is the average value of $f(\rho_r)$'s;
and the standard deviation ($\delta f$), which stands for 
the random error. 

On one hand, the  standard deviation ($\delta f$)
obtained from type-I
is $\pm 0.115$, while those ones for type-II($\theta$)
with $\theta = 0.5\pi$ and $\theta = 0.1\pi$ are $\pm 0.146$
and $\pm 0.208$, respectively. 
[Note that hereafter we add a suffix $(\theta)$ to ``type-II"
to declare the $\theta$-dependence of this type.]
Among these cases, the random error of 
type-I is smallest while that value of type-II
gradually increases when reducing the 
interaction strength $\theta$. 
These results are in agreement with 
\cite{PhysRevLett.116.040502}.

On the other hand, the average fidelities 
obtained from these three cases above are
$0.852, 0.837,$ and $0.718$, respectively. 
These average fidelities are smaller 
and deviate systematically from the true value (0.9),
which are known as the systematic errors 
\cite{PhysRevLett.114.080403}.
To evaluate a `` measure" of systematic error, 
we define a bias factor,  
which is the ratio between the reference fidelity $f_0$
and the evarage fidelity $f_{\rm ave}$ as
\begin{align}\label{eq:bias}
\Delta f = \dfrac{f_0-f_{\rm ave}}{f_0},
\end{align}
 where $f_0-f_{\rm ave}$ is an infidelity. 
If the bias vanishes ($\Delta f=0$), 
there is no systematic error at all.
The larger the bias, the larger the systematic error.
The analytical expression of \eqref{eq:bias}
can be derived in terms of the initial unknown
state. However, we emphasize that such an expression
is given in an ideal case, such as the number of measurements is infinity. In this paper, we thus restrict ourselves to numerical simulation of \eqref{eq:bias}, which can be served as
testbed before carrying out real experiments. 

As we can see from Fig.~\ref{fig2}, 
the bias is 
small for type-I, 
while it gradually increases for type-II($\theta$)
when $\theta$ gradually reduces from strong to weak interactions.
These results are obtained after 400 measurement 
runs ($N_c = 400$) for each $\rho_{nm}$
and 500 estimated states to get the histogram. 

%
\begin{figure*} [t]
\centering
\includegraphics[width=16cm]{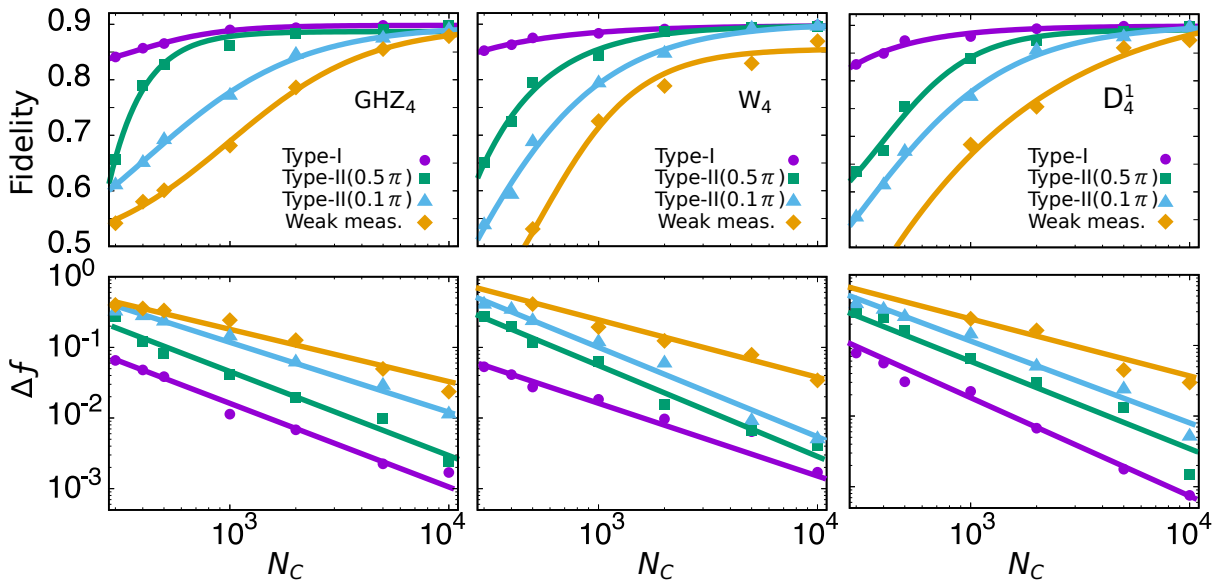} 
\caption{
(Color online.) Upper panels: 
The average fidelities with respect to the target state as functions of 
the number of copies $N_c$ for different target states 
GHZ$_4$, W$_4$, and D$_4^1$ from left to right. 
For each panel, we consider different methods, 
ranging from type-I to type-II($\theta$) and weak measurements.
Lower panels: the corresponding $\Delta f$'s.
The solid lines are the guide's eyes.
}
\label{fig3}
\end{figure*} 

While random errors can be reduced
when increasing the number of copies $N_c$, 
systematic errors cannot be completely eliminated in 
the same way.
However, the bias $\Delta f$ can be reduced 
when increasing $N_c$, for example, see
\cite{PhysRevLett.114.080403};
 or in a general procedure called ``calibration."
Hereafter, let us investigate the dependence of fidelities on $N_c$.
The results are shown in the upper row of Fig.~\ref{fig3}
for several initial states, including the GHZ state $|{\rm GHZ_4}\rangle$, 
the W state $|{\rm W_4}\rangle$, and the Dicke state $|{\rm D_4^1}\rangle$.
For each case, we show the average fidelities 
with respect to the target state for 
type-I (purple), type-II($0.5\pi$)(green),
type-II($0.1\pi$) (cyan), 
and weak measurement (orange).
The results show that these fidelities depend significantly on 
the number of measurement copies $N_c$ 
and increase with increasing $N_c$.  
Correspondingly, the biases $\Delta f$ decrease with increasing $N_c$, 
as can be seen from the lower panels of Fig.~\ref{fig3}. 
The results also suggest that the biases still remain and 
do not collapse between deferent methods when increasing $N_c$,
which imply that the systematic errors cannot be completely eliminated.
It can be seen that the systematic error is small for 
type-I, and it gradually increases for type-II($\theta$) 
when $\theta$ reduces from 
strong to weak interactions. 
We also compare to the weak measurement case,
where the interaction strength is infinitely small. 
The results reveal that the systematic error in this case
is the largest (among those methods that we are considering 
in this paper.) This conclusion is in agreement with 
previous studies 
\cite{PhysRevLett.116.040502,PhysRevA.89.022122},
where the bias caused by the infinitely small $\theta$.

\begin{figure*} [t]
\centering
\includegraphics[width=16cm]{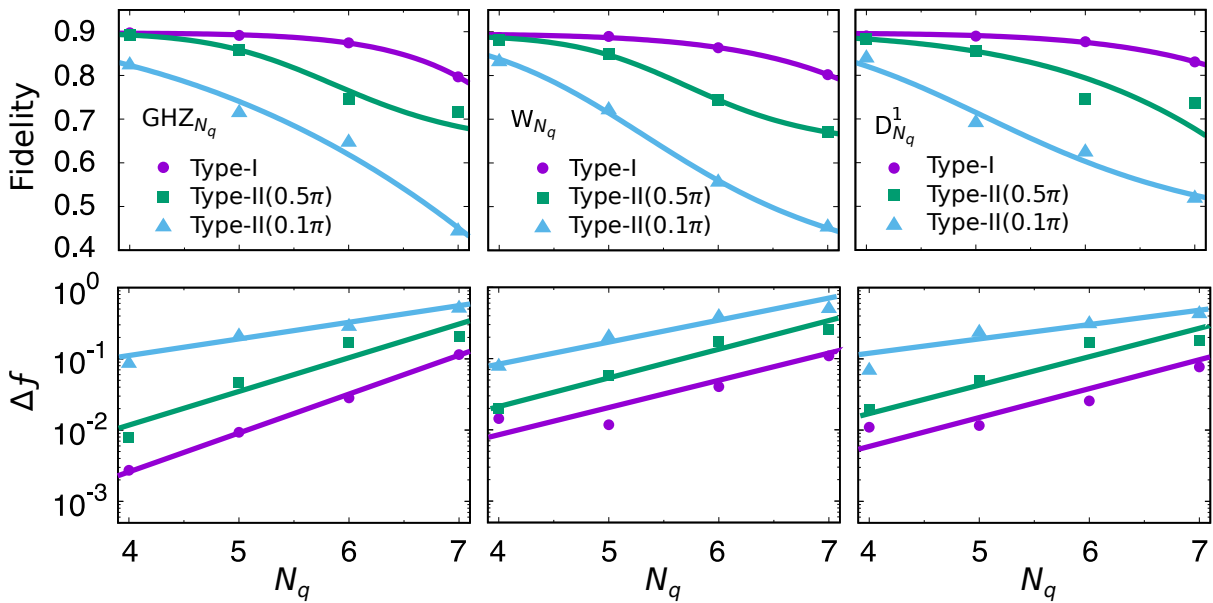} 
\caption{
(Color online.) Upper panels: 
The average fidelities with respect to the target state as functions of 
the number of qubits $N_q$ for different target states 
GHZ$_4$, W$_4$, and D$_4^1$.
Other denotations are the same as Fig.~\ref{fig3}. 
Lower panels: the corresponding $\Delta f$'s.
}
\label{fig4}
\end{figure*}

Furthermore, the systematic errors also vary with 
the number of qubits $N_q$.
In Fig.~\ref{fig4} upper panels, 
we show the average fidelities with respect to the target state
of different methods versus 
$N_q$ for different target states.
We also show the corresponding 
biases in its lower panels.  
Here, we fit $N_c = 10^5$.
The results suggest that the fidelities decrease  
when increasing $N_q$ while the biases are increasing,
which imply the increase in systematic errors. 
Similarly as above, for type-I,
the systematic deviation is small while it gradually increases 
for type-II($\theta$) when $\theta$ reduces from 
strong to weak interactions.
Here we omit the weak measurement case since 
its systematic deviation is relatively large.

\subsection{Confidence region}
Now, let us analyze the confidence region in the DSM. 
Following the method of Christandl and Renner 
\cite{PhysRevLett.109.120403},
we consider a region $R_{f(\rho_r)}$ is a set of
state $\rho_r$ so that its fidelity $f(\rho_r)$ satisfies a concrete 
condition
\begin{align}\label{eq:Rf}
R_{f(\rho_r)} =\bigl\{\rho_r: \bar{f} \le  f(\rho_r) \le 2f_0-\bar{f} \le 1\bigr\},
\end{align}
where $\bar{f}$ is a fixed threshold fidelity. 
The distribution function of 
$f$ for all $\rho_r$ is also defined as
\cite{PhysRevLett.117.010404}
\begin{align}\label{eq:dist}
\mu(f) = \int_{\forall\rho_r} d\rho_r\ P(\rho_r)\delta[f(\rho_r)-f],
\end{align}
where $P(\rho_r)$ is the normalized probability that the 
obtained state is $\rho_r$, and $\delta[\cdot]$ is the Dirac delta.
In the region $R_{f(\rho_r)}$, the distribution function is satisfied 
\cite{PhysRevLett.109.120403,PhysRevLett.117.010404}:
\begin{align}\label{eq:alpha}
 \int_{\bar{f}}^{{\rm min}\{2f_0-\bar{f},1\}}\mu(f) df
\ge 1-\dfrac{\epsilon}{2c},
\end{align}
where $1-\epsilon$ is the confidence level
and $c \equiv \text{poly}(N_c) = (N_c+1)^{d-1}$.
Then, there exists a confidence region $R_{f(\rho_r)}^\lambda$
such that
\begin{align}\label{eq:Rfs}
R_{f(\rho_r)}^\lambda =\bigl\{\rho_r: \exists \rho'_r \in R_{f(\rho_r)} 
\text{ with } F(\rho_r,\rho'_r) \ge 1-\lambda^2\bigr\},
\end{align}
where $F(\rho_r,\rho'_r)  = {\rm tr}
\sqrt{\sqrt{\rho_r}\rho'_r\sqrt{\rho_r}}$ the fidelity
of the closeness between $\rho_r$ and $\rho'_r$, 
and 
%
$\lambda^2 = \dfrac{2}{N_c}(\ln\frac{2}{\epsilon} + 2\ln c)$.
%
Then, the probability that $\rho_r$ 
belongs to the confidence region 
is given by
\begin{align}\label{eq:pro}
P\bigl[\rho_r \in R_{f(\rho_r)}^\lambda\bigr] \ge 1-\epsilon.
\end{align}

\begin{figure*} [t]
\centering
\includegraphics[width=16cm]{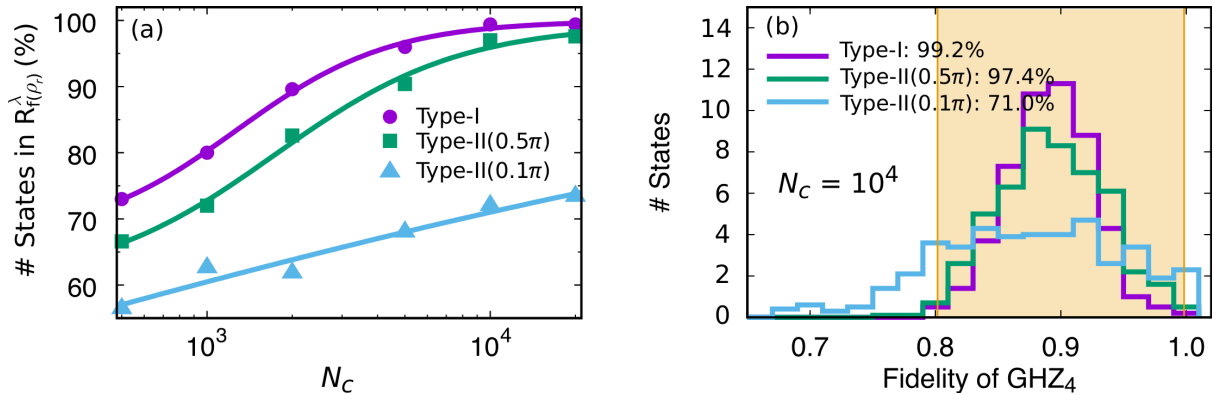} 
\caption{
(Color online.) (a)
Plot of the ratio into percentage between 
the number of states inside the 
confidence region and all states. 
We plot for three cases shown in the figure. 
(b) The histogram of these three cases at $N_c = 10^4$.
The ratios into percentages are given in the figure, 
while the highlight region represents the confidence region.
}
\label{fig5}
\end{figure*}

Next, let us provide a concrete example of how to choose
the confidence region priorly. 
Let us assume the distribution function $\mu(f)$ is a Gaussian
mean $f_0$:
\begin{align}\label{eq:muf_Gau}
\mu(f) = \dfrac{1}{\sqrt{2\pi\sigma^2}}e^{-\frac{(f-f_0)^2}{2\sigma^2}},
\end{align}
where we choose priorly $\sigma = \epsilon = 0.005$, i.e., 
the confidence level is $1-0.005 = 0.995$.
Inversely solving \eqref{eq:alpha} we obtain 
$\bar{f} \approx 0.858128$. 
We also calculate $\lambda^2$ and then obtain the confidence region
bounded by 
\begin{align}\label{eq:conf}
\bar{f} - \lambda^2\le R_{f(\rho_r)}^\lambda\le
 \min(2f_0-\bar{f}+\lambda^2,1).
\end{align}
For example, with GHZ$_4$, $N_c = 10^4$, we have 
$\lambda^2 \approx 0.056461$, then the
confidence region is $[0.801667: 0.998333]$,
which is the yellow region in Fig.~\ref{fig5}(b).

In Fig.~\ref{fig5} (a), we numerically show 
the ratio into percentage between 
the number of states $\rho_r$ 
that belong to the confidence region 
$R_{f(\rho_r)}^\lambda$ and all states
with the confidence level at 0.995.
We investigate three cases of type-I,
type-II(0.5$\pi$)
and type-II($0.1\pi$) for several $N_c$.
The results show that the ratio into percentage
increases as $N_c$ increases; 
and type-I always shows 
the highest percent while that one for type-II gradually decreases 
from strong to weak interactions.   

In Fig.~\ref{fig5} (b), we show the histogram of these three cases
where we can see that type-I lies inside the confidence region
with the highest percentage. 

\subsection{Systematic errors against the noise} 
In this subsection, we investigate the systematic errors
under the  effect of noise caused by 
the imperfect detection. 
In this case, the  
probability when measuring the control qubit probe
($P_j$)
has some noises. We consider the noise a Gaussian type 
\cite{PhysRevLett.110.163604,PhysRevLett.115.163002},
then the probability is given as 
\begin{align}\label{eq:Gauss}
P_j^{(\eta)} = \sum_{j'} 
\mathcal{N}e^{-\frac{(j-j')^2}{2\eta^2}} P_{j'},
\end{align}
where $\mathcal{N}$ is the normalizing factor, 
and $\eta$ is the noise parameter.
Physically, the imperfect detection noise means that 
when measuring the control qubit probe in basic $|j\rangle$, there is a 
small probability that it becomes $|j'\rangle$ for
$j$ and $j'$ are in $\{ |0\rangle, |1\rangle\}$, 
$\{|+\rangle, |-\rangle\}$, and $\{|L\rangle, |R\rangle\}$.

\begin{figure} [t]
\centering
\includegraphics[width=8.2cm]{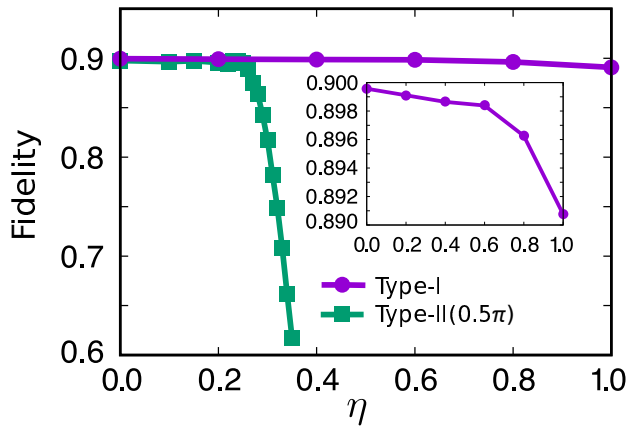} 
\caption{
(Color online.) Plot of average fidelities of GHZ$_4$ at $N_c = 10^4$
for two cases:
type-I and type-II($0.5\pi$). 
Inset: plot of the fidelity of type-I
on a large scale. 
}
\label{fig6}
\end{figure}

We show, in Fig.~\ref{fig6}, the numerical results for the average fidelity of 
GHZ$_4$ for two cases of type-I
and type-II ($0.5\pi$). 
It can be seen that the fidelity is protected 
in the case of type-I,  which implies that
its systematic error does not increase when increasing the noise.
Meanwhile, the fidelity of type-II($0.5\pi$) 
rapidly drops when the noise $\eta$ reaches around 0.3.

\section{Discussion}\label{sec_iv}

In our study, we show a 
higher accuracy for type-I operational interaction
in comparison with type-II from strong to weak interactions. 
It can be understood because in type-I,
we only measure $\rho''_{10}(n,k)$ [equation \eqref{eq:rho10_I}] 
while in the latter case, 
it also requires to measure 
$\rho''_{11}(n,k)$ [equations (\ref{eq:rho10_II},\ref{eq:rho11_II})]. 
Consequently, type-I gives more accuracy. 
Both cases of type-I and weak measurement 
only require the measurement of $\rho''_{10}(n,k)$.
However, for the weak measurement case, the 
bias is rising due to the infinitely small 
of the interaction strength. 
As a result, the accuracy of the weak measurement case
is poor. 

So far, the quantum controlled interaction presents 
a cyclic property between pre-selected state, basis $|n\rangle$,
and post-selected state $|k\rangle$, 
which allows for applying the scan-free method \cite{Ogawa_2019}. 
In the scan-free method, the projection operator 
$|n\rangle\langle n|$ in \eqref{eq:U_B} and \eqref{eq:U_B_II}
is replaced by 
$|k\rangle\langle k|$, 
and the postselection is $|n\rangle$.
We can keep all the data of the postselection $|n\rangle$
(scan-free) that will be used in the reconstruction process.
It thus gives better accuracy for the DSM \cite{Shi:15,Ogawa_2019}.

Furthermore, the study of confidence region 
is crucially important not only in theory but also 
in experiments because it helps to design and 
evaluate experiments.  For example, it allows 
experimenters to predict the confidence region 
of quantum states in advance before carrying 
out the experiments. Thus, the measurement time 
can be reduced by focusing on the confidence region.
 So far, the analysis of the confidence region 
has been carried out for the standard 
quantum state tomography
\cite{PhysRevLett.109.120403,PhysRevLett.117.010404,PhysRevLett.122.190401,Shang_2013}. 
However,
this is the first time we apply this analysis on the DSM.
\sloppy
\section{Conclusion}{\label{sec_v}
We have investigated the systematic 
operational errors in the direct state measurements (DSM)
with a quantum controlled interaction framework.
In this DSM scheme, a  target system is
controlled by a qubit probe and postselected onto a conjugate basis;
the outcomes of the control qubit probe can be measured 
and taken out of the estimated state. 
We have considered two types of operational interaction: 
(i) invert quantum controlled interaction (type-I) 
and (ii) arbitrary strong interaction (type-II), 
which is equivalent to a von Neumann interaction.

Our numerical results first show that 
type-I operational interaction gives lower 
systematic error than type-II, which means more accuracy.
For the same confidence level, 
type-I has a higher ratio into percentage in the confidence region.
The systematic error in type-I is also well against the noise
as its fidelity is protected under the noise.
These results can be explained by the difference
in these two interaction types, where type-I requires
fewer measurements than type-II.

Our study gives a better solution for 
quantum state tomography using the quantum
controlled measurement scheme 
(better than conventional strong measurements and weak measurements.)
Furthermore, its measurement is simple
and applicable to high-dimension systems.
This method, thus, could be a potential candidate for 
further characterizing the properties of large systems. 

\section{Acknowledgments}
We acknowledge Y. Kondo for the careful reading of the manuscript. This work was supported by JSPS KAKENHI Grant Number 19K14620. 

\bibliographystyle{apsrev4-1}
\bibliography{mybib}

\end{document}